\documentclass[]{spie}  

\usepackage{amsmath,amsfonts,amssymb}
\usepackage{graphicx}
\usepackage[colorlinks=true, allcolors=blue]{hyperref}

\title{PSF nowcast using PASSATA simulations - Towards a PSF forecast}

\author[a]{Turchi, A.}
\author[a]{Agapito, A.}
\author[a]{Masciadri, E.}
\author[b]{Beltramo-Martin, O.}
\author[d]{Milli, J.}
\author[a]{Plantet, C.}
\author[a]{Rossi, F.}
\author[a]{Pinna, E.}
\author[c]{Sauvage, J.F.}
\author[b]{Neichel, B.}
\author[c]{Fusco, T.}
\affil[a]{INAF-Osservatorio Astrofisico di Arcetri, L.go Enrico Fermi 5, Firenze, Italy}
\affil[b]{LAM-Laboratoire d’Astrophysique de Marseille, UMR 7326, 13388, Marseille, France}
\affil[c]{ONERA, B.P. 72, F-92322 Chatillon, France}
\affil[d]{IPAG-Institut de Planétologie et d’Astrophysique de Grenoble, 414, Rue de la Piscine, 38400
St-Martin d’Hères, France}

\authorinfo{Send correspondence to Alessio Turchi - e-mail: alessio.turchi@inaf.it}

\begin{document} 
\maketitle

\begin{abstract}
Characterizing the PSF of adaptive optics instruments is of paramount importance both for instrument design and observation planning/optimization. Simulation software, such as PASSATA, have been successfully utilized for PSF characterization in instrument design, which make use of standardized atmospheric turbulence profiles to produce PSFs that represent the typical instrument performance. In this contribution we study the feasibility of using such tool for nowcast application (present-time forecast), such as the characterization of an on-sky measured PSF in real observations. Specifically we will analyze the performance of the simulation software in characterizing the real-time PSF of two different state-of-the-art SCAO adaptive optics instruments: SOUL at the LBT, and SAXO at the VLT. The study will make use of on-sky measurements of the atmospheric turbulence and compare the results of the simulations to the measured PSF figures of merit (namely the FHWM and the Strehl Ratio) retrieved from the instrument telemetry in real observations. Our main goal in this phase is to quantify the level of uncertainly of the AO simulations in reproducing real on-sky observed PSFs with an end-to-end code (PASSATA). In a successive phase we intend to use a faster analytical code (TIPTOP). This work is part of a wider study which aims to use simulation tools joint to atmospheric turbulence forecasts performed nightly to forecast in advance the PSF and support science operations of ground-based telescopes facilities. The ‘PSF forecast’ option might therefore be added to ALTA Center or the operational forecast system that will be implemented soon at ESO.
\end{abstract}

\keywords{simulations,adaptive optics,forecast,atmosphere,optical turbulence}

\section{INTRODUCTION}
\label{sec:intro}

Atmospheric optical turbulence (OT) is the main cause for the degradation in performance of the observations from ground-based telescopes. The deformation induced by OT on the incoming wavefront from space can be roughly summarized in the spreading of the point spread function (PSF), which describes the response of the telescope optics to a point light source. The PSF quality is influenced by many parameters, however a detailed description of these effects is out of the scopes of the present paper. To get a summarized description. A top-class telescope with an 8-meters diameter (D) primary mirror, in theory, could achieve an angular resolution $\sim \lambda/D$, which equals to 0.13" in V band. The OT is typically measured with the seeing ($\epsilon$) parameter, which, on the best observing sites, has a typical median value around 0.7" in V band (order of magnitude). The seeing represents the spreading of the PSF and its value sets a hard limit to the angular resolution achievable from ground. This conditions is typically referred as "seeing limited", and the main limiting factor at least until infrared bands (Q).\\
The Adaptive Optics (AO) aims to overcome this limitation and correct the deformations induced on the weavefront by the OT. However the AO efficiency is affected by the optical turbulence itself.\\
Most AO facilities may use different operation modes that are able to perform different kind of OT corrections, depending on the needs of the scientific programs and the available atmospheric conditions. Single Conjugated Adaptive Optics (SCAO) aims to obtain the best performances over a limited portion of the sky. If the target is near to the guide star (on-axis case), the shape of the PSF depends only on the integrated turbulence parameters (i.e. seeing, $\tau_0$...), however in the off-axis case, where the guide star is separated from the target, the PSF is impacted by the vertical structure of the turbulence ($C_N^2$ profile) and the wind speed. This is even more important for the Wide Field AO case (WFAO), where the correction is applied to a large portion of the sky. In any case the knowledge of the PSF attainable in specific conditions is fundamental to correctly match the AO facility and the scientific program to the best atmospheric conditions.\\
The knowledge (in advance) of the OT can finally help in planning telescope observations and tuning the AO instruments for better performance [\citenum{masciadri2013}, \citenum{masciadri2020}].\\

Knowledge in advance of the atmospheric turbulence is possible only with a forecast model, which is also able to provide the full $C_N^2$ stratification. Since such topic is of paramount importance and have such a deep impact on ground-based observations, in recent years many telescopes, with top-class AO facilities, expressed interest in having a forecast tool for OT. Specifically, the usage of dedicated atmospheric models to forecast the atmospheric turbulence had a significant growth in the last decade, with more and more large telescope installations adopting such tools. Mauna Kea (KECK telescope) already had a forecast system for OT\footnote{\href{http://mkwc.ifa.hawaii.edu/current/seeing/}{http://mkwc.ifa.hawaii.edu/}} . The Large Binocular Telescope (LBT) followed with the ALTA project\footnote{\href{http://alta.arcetri.inaf.it/}{http://alta.arcetri.inaf.it/}} . The latter uses an advanced dedicated system for forecasting OT using a high-accuracy mesoscale atmospheric model (Astro-Meso-NH, [\citenum{masciadri1999}]), which has been proven to provide reliable forecasts of atmospheric parameters [\citenum{turchi2017}] as well as the seeing [\citenum{masciadri2020}] above the LBT site. Now also the ESO's Very Large Telescope (VLT) is planning to adopt a similar system during the course of next year which could potentially have a huge impact on the astronomy community due to the importance of the AO facilities at VLT and the huge scientific output of the telescope [\citenum{masciadri2017,masciadri2022}].\\
Recently it has been proven that, using an autoregression technique, it is possible to forecast the seeing with performances (RMSE) at short time scale (1h) of the order of 0.1" [\citenum{masciadri2020,masciadri2022}] . Such a value is better than the intrinsic uncertainty of measurements and this guarantees us to be able to achieve forecast of the optical turbulence with a sufficiently good accuracy.\\

In this paper we want to dedicate our attention on the possibility to forecast a few features of the PSF obtained by AO systems. AO performance depends mainly on the OT and atmospheric conditions, the characteristics of the observed target (magnitude M) and on the fine tuning of the optical and control-loop parameters to the matching conditions. Each AO instrument is different and have different specifications, however by joining in succession, a forecast system of the atmospheric and turbulence conditions with an AO simulation software that is able to simulate the performance of the optics and the control loop of the AO system, it is conceivable to provide a prediction of the AO performance for each specific observed target and atmospheric condition. This could provide an invaluable tool for the telescope operator that could carefully plan the observation to match the best possible conditions and achieve the level of performance required, thus boosting the scientific output of the telescope.\\
In the present paper we will limit ourselves to the study of the Full Width at Half Maximum (FWHM) and Strehl Ration (SR), which are the principal numbers that characterize the goodness of the PSF corrected by the AO and the attainable performance on an observed target. We might envisage in the future to include other parameters such as the encircle energy or the contrast.\\

In the present paper we will provide an update to the results previously shown in [\citenum{turchi2020s}], in order to characterize and potentially evaluate the feasibility of such a forecast system for FWHM and SR for few of the best AO systems currently available. SOUL at the LBT and SAXO (the AO instrument working with the planet finder SPHERE) at the VLT. OT predictions can also provide detailed high-accuracy profiles for the  stratification of the turbulence ($C_N^2$ profiles), which could have a huge impact for off-axis predictions. However, since this is a preliminary feasibility study, in the present paper, we will limit ourselves to the on-axis case which can be typically characterized using only the integrated seeing parameter (the seeing is the integral of the $C_N^2$ along the line of sight).\\

Our interest is also to compare different AO simulation software in order to assess the best tool possible for real-time forecast applications, which have the requirement of fast computation times in order to provide a timely forecast.\\

In section \ref{sec:atmo} we will resume the atmospheric model configuration with the best attainable OT forecast performance using short-term prediction strategies.\\
In section \ref{sec:aosim} we will briefly present the AO simulation software used for the present study.\\
In section \ref{sec:soul} we will show how the current state-of-the-art of FWHM and SR predictions using the best possible OT knowledge.\\
In section \ref{sec:tiptop} we will show a comparison of different AO simulations software and a preliminary feasibility study of how one could setup a real-time operational forecast system, to be used as a forecast service for the telescope and the AO facilities.\\
In section \ref{sec:final} we will draw the conclusions and resume the results shown in the present paper.

\section{Atmospheric model}
\label{sec:atmo}

In this section we describe the specifications of the atmospheric model with which we intend to provide the forecast of the seeing. The seeing is therefore the input of the AO software.

We intend to use the Astro-Meso-NH package [\citenum{masciadri1999}] to forecast the atmospheric OT over the studies sites, LBT and VLT. The Astro-Meso-Nh code is a dedicated package conceived to predict the OT. It is based on the use of an atmospheric model called Meso-NH [\citenum{lafore98,lac2018}].  We refer to Turchi et al. [\citenum{turchi2017}] for more details of the numerical configuration for the LBT site (ALTA project). For the VLT case we refer to  Masciadri et al. [\citenum{masciadri2013}]. Initial conditions (used for model initialization and periodical forcing of external conditions) are provided by the General Circulation Model (GCM) run by the European Center for Medium Weather Forecasts (ECMWF), with an horizontal resolution of 9 km.\\
The long-term forecast produced by the model is typically made available during the afternoon before the start of the night. In recent years the performances of OT atmospheric prediction models increased dramatically (e.g. [\citenum{masciadri2020}]). Specifically we see that the trend in OT prediction is to move from long-term predictions, which typically provide knowledge of the OT conditions from few hours to few days before the start of the observing night, to short-term predictions, which provide OT predictions few (1-4) hours in advance from the planned observing time.\\ This method improves the accuracy of the predictions over the next few hours in the future and allows to maximize the accuracy and still gives precious information for observation planning. This short-term forecast is produced each hour during the night and allows extreme forecast accuracy, up to a RMSE of 0.1 arcsec with respect to the seeing parameter at an hour in the future [\citenum{masciadri2020}].\\

We remember that the typical uncertainty on the seeing measurement obtained with different instrument is of order $\sim0.2"$ (see \citenum{osborn2018}). This means that we already have a seeing prediction that is above the conceivable accuracy of an instrument, being practically indistinguishable from the measurement itself through an instrument validation.\\

A system similar to ALTA is currently under development for ESO at VLT and it has already been possible to prove its abilities in providing PWV forecast at very high accuracy [\citenum{turchi2020}] at short time scales. Performances obtained for the OT are comparable to what observed at the LBT site or even better \cite{masciadri2022}.

In this contribution we are interested in trying to evaluate the best possible performance obtainable for the PSF forecast in order to define the specifications for the realization of an operational PSF forecast system. We already know which are the performances of the forecast of the OT and the accuracy we can achieve. We therefore concentrate on the performances of the AO simulation software to evaluate which accuracy we can achieve in this second step. 


\section{AO simulation software}
\label{sec:aosim}

In this work we consider two AO simulations software: PASSATA [\citenum{agapito2016}] and TIPTOP [\citenum{neichel2021}]. PASSATA is an end-to-end simulator while TIPTOP is a Fourier based simulator. PASSATA was used during SOUL design and commissioning and we are confident in its ability to  accurately estimate SOUL performance  [\citenum{pinna2019}].
TIPTOP is a more recent tool that has some indisputable advantages in terms of computational time (see discussion in [\citenum{turchi2020s}]) and it showed a good capability of fitting on sky images [\citenum{pinna2019,kuznetsov2022}].
Please note that both simulators produce noiseless and infinite dynamic range PSFs so they are not affected by errors in the computation of SR and FWHM as the real images.
We decided to include most of the error terms in the simulations except residual non common path aberrations (NCPA, residual because we are not interested in the absolute value of NCPA, but only in its residual after correction [\citenum{esposito2020}]) and we consider vibrations only for SOUL case where we have a good knowledge of the vibration level [\citenum{Kulcsar2012,agapito2021}].
We remind that in principle there is no interest in adding residual NCPA and vibrations in the simulations, but they can be added a posteriori scaling the SR with the Marechal equation.
In fact, we cannot forecast the vibration level nor the residual NCPA, they are errors that must be produced by some statistical analyses of the instrument.
In particular, as we reported in [\citenum{agapito2021}], vibration level can change quickly and can have a wide range of values.
In theory there could be some correlation with the atmospheric parameters, for example wind direction and speed, but it will also depend on the orientation of the dome, by the elevation angle and by possible wind shields.
As we will show in the following vibrations pose a limit to the accuracy of the performance estimation, in particular on the FWHM that is more sensitive to them.

\section{AO simulations on SOUL measurements}
\label{sec:soul}

\begin{figure}
\centering
\includegraphics[width=1.\textwidth]{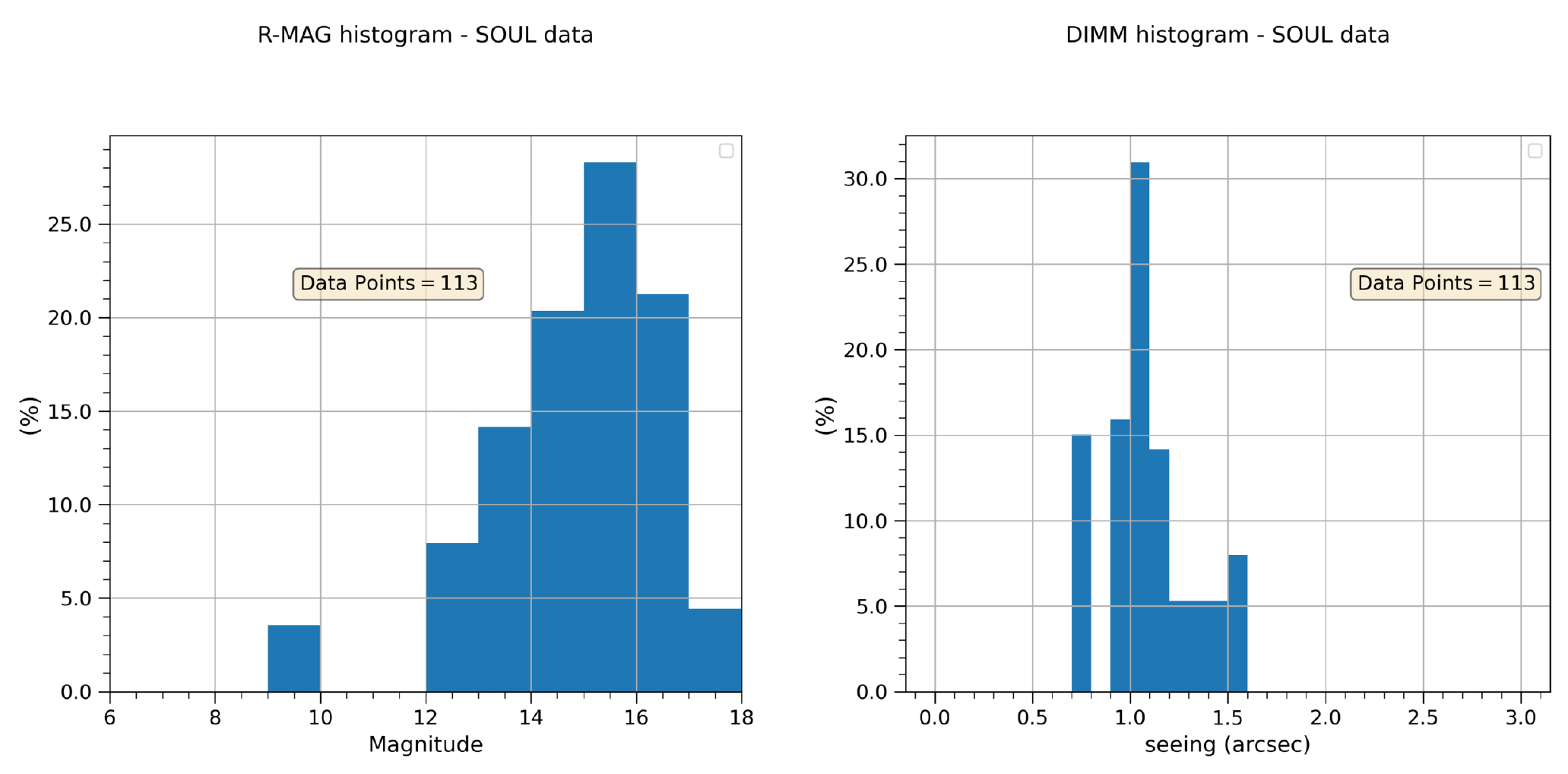}
\caption{Statistics of the SOUL measurement samples. Left: Distribution of the magnitude of the observed stars in R band (used for simulations). Right: Distriution of the seeing values as measured by DIMM.}
\label{fig:soulmagplot}
\end{figure}

In order to evaluate the possible performances of a PSF prediction, we used a set of PSF measurements provided by the AO team responsible for SOUL commissioning and stored in the AO telemetry. SOUL [\citenum{pinna2019}] is an extreme single conjugate AO system that feeds the camera LUCI at LBT, and is the upgrade of the previous FLAO system. We stress that, at present, we do not have an estimate for the accuracy of SR and FWHM measurements estimated from the LUCI images. In this section we show the PSF predictions obtained using the PASSATA software [\citenum{pinna2016}], using, as an input, the observed magnitude in R-band (the one used by the Wafefront Sensor (WFS)) and the seeing provided by the DIMM measurements. We already shown that the forecast error of the seeing is smaller than the uncertainty obtained with estimates from different instruments. This tells us that using the DIMM directly is more than adequate for this study. In practice we are expressly considering only the uncertainty related to the AO simulation tool without taking into account the uncertainty coming from the atmospheric forecast system to be able to control the AO software performances.\\
We had a total of 113 PSF measurements taken on 18 individual nights, both in H and Ks bands. SOUL operates efficiently on very faint magnitudes, so most of the targets considered in this measurement sample falls in the range of magnitude $[12-18]$, as shown in Fig. \ref{fig:soulmagplot}. In the same figure we also show the distribution of the seeing observed in this sample, as measured by the DIMM, showing values which are typical for LBT (median seeing of the telescope is around $\sim1"$). We remember that in the LBT case the DIMM is placed inside the dome, so on one hand it tends to overestimate the purely atmospheric seeing with additional effects from the dome seeing, however on the other hand, since the short-term prediction is autoregression based (thus uses instrument measurements), it allows us to consider also the dome seeing in the OT forecast itself, thus providing a more accurate estimate of the real seeing impacting on the telescope optics.\\

\begin{figure}
\centering
\includegraphics[width=1.\textwidth]{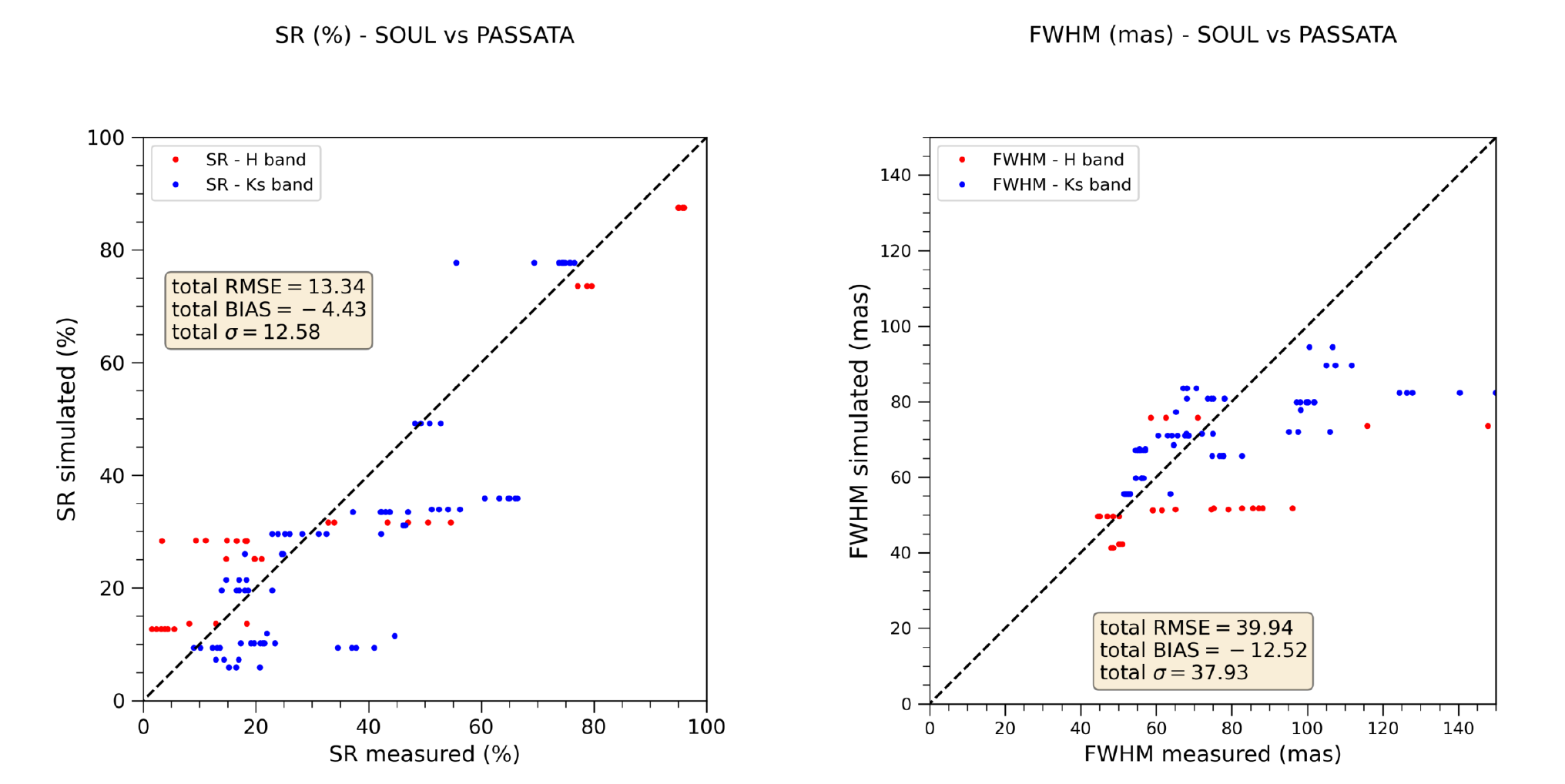}
\caption{Scatterplot of the results of the simulations with PASSATA for the SOUL case. Left: comparison of simulated and measured SR. Right: Comparison of simulated and measured FWHM. In each figure we report the total RMSE, BIAS and $\sigma$ obtained for the whole sample, while with different colors we plot the two observed bands (H and Ks).}
\label{fig:soulscatterplot}
\end{figure}

\begin{figure}
\centering
\includegraphics[width=1.\textwidth]{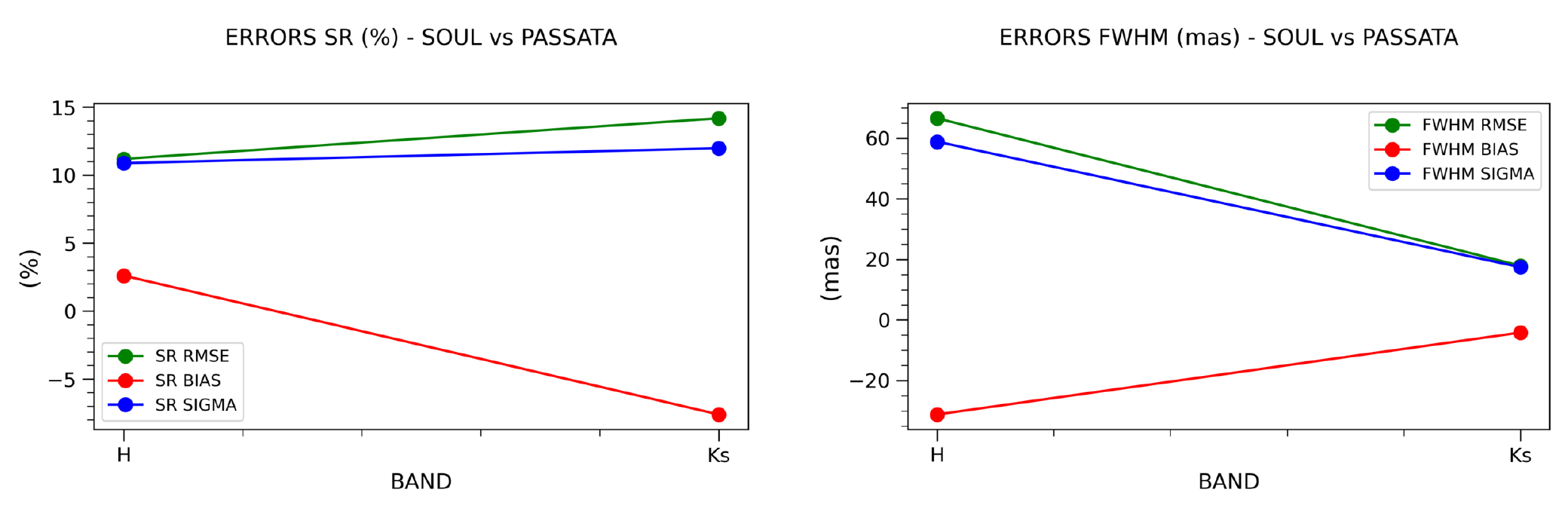}
\caption{Errors reported for each observed band in the SOUL simulations with PASSATA. We report the RMSE, the BIAS and the SIGMA. Left: Errors on SR. Right: Errors on FWHM.}
\label{fig:soulerrorplot}
\end{figure}

The PASSATA simulation strategy used in the SOUL case consists in performing 3 different short (a few seconds of simulated time) simulations runs, each with a different random atmospheric and noise realization, and averaging the 3 results in order to produce a statistically reliable estimate. The simulation length has been optimized to allow convergence of the simulation results. The AO system parameters (time\_step, detector binning, number of modes, gain, ttgaim, mod\_amp) are selected in the same way of the real system as a function of the guide star magnitude [\citenum{pinna2019}]. Specifically the parameters are optimized in the following magnitude ranges [0-13], [13-14.5], [14.5-16.5] and [16.5-18.5]. Each simulation run performs two different set of simulations for the extremes of one of the above magnitude ranges where the observed magnitude falls in. Then we obtain the final result, with the observed magnitude, with a linear fit from the two obtained values in the selected magnitude range. Please note that this strategy has been designed to cope with the time requirement of a forecast system working with a short advance.
The results, in terms of predicted SR and FWHM, are shown in Fig. \ref{fig:soulscatterplot} and \ref{fig:soulerrorplot}, the latter with the RMSE, bias and sigma reported for each observed band.\\

In Fig. \ref{fig:soulscatterplot} we see that most of the measurement, those taken within a short time interval between each other (seconds), show an horizontal distribution that we imputed to the impact of the telescope vibrations and windshake, which are the main effect contributing to the noise on such short-term intervals (see also considerations in Sec.\ref{sec:aosim}). The correlation between prediction and measurements is however evident from the scatterplot, with a global BIAS of order $\sim4\%$, while the RMSE varies in the range [10-15\%] from H to Ks band. This in extremely encouraging result which show that, for short-term  predictions, the SR forecast is already at a level that could provide extremely accurate information to the telescope operator.\\

The FWHM forecast has a larger dispersion, despite the correlation being evident, however this is especially true only in the H band (with an RMSE $\sim60 mas$). In this case we see the large effect of vibrations and windshake that tend to disperse the temporally correlated measurements over a wide horizontal range. In the Ks band the error is of order $\sim 20 mas$, with a negligible bias, probably also thanks t a larger statistics (see Fig. \ref{fig:soulerrorplot}.\\

\section{Comparison between AO simulation software}
\label{sec:tiptop}

PSF real-time forecast, if using  short-term seeing predictions, requires a fast computational time in order to provide a timely forecast that can be of any help to the telescope planning. PASSATA, while being GPU accelerated, is relatively slow since it's an end-to-end full simulation software, requiring several minutes for each simulation. Also it is more complex to transform in an operational tool, able to be run 24/7 in an automated way, due to several technical reasons that we will not discuss in detail in this paper. While tecnhically feasible, an operational setup that can match the simulations times required for an operational tool is relatively expensive, needing one or more dedicated GPU servers.\\

In order to address the previous issue, we started evaluating a novel AO simulation software, TIPTOP [\citenum{neichel2021}], which has the benefit of being an analytical tool, thus providing much faster computation times with much less computational power, leading to less expensive required hardware and more simple operative setups. Specifically, results are obtained in seconds on a consumer-level desktop PC, practically with negligible delays.\\

\begin{figure}
\centering
\includegraphics[width=1.\textwidth]{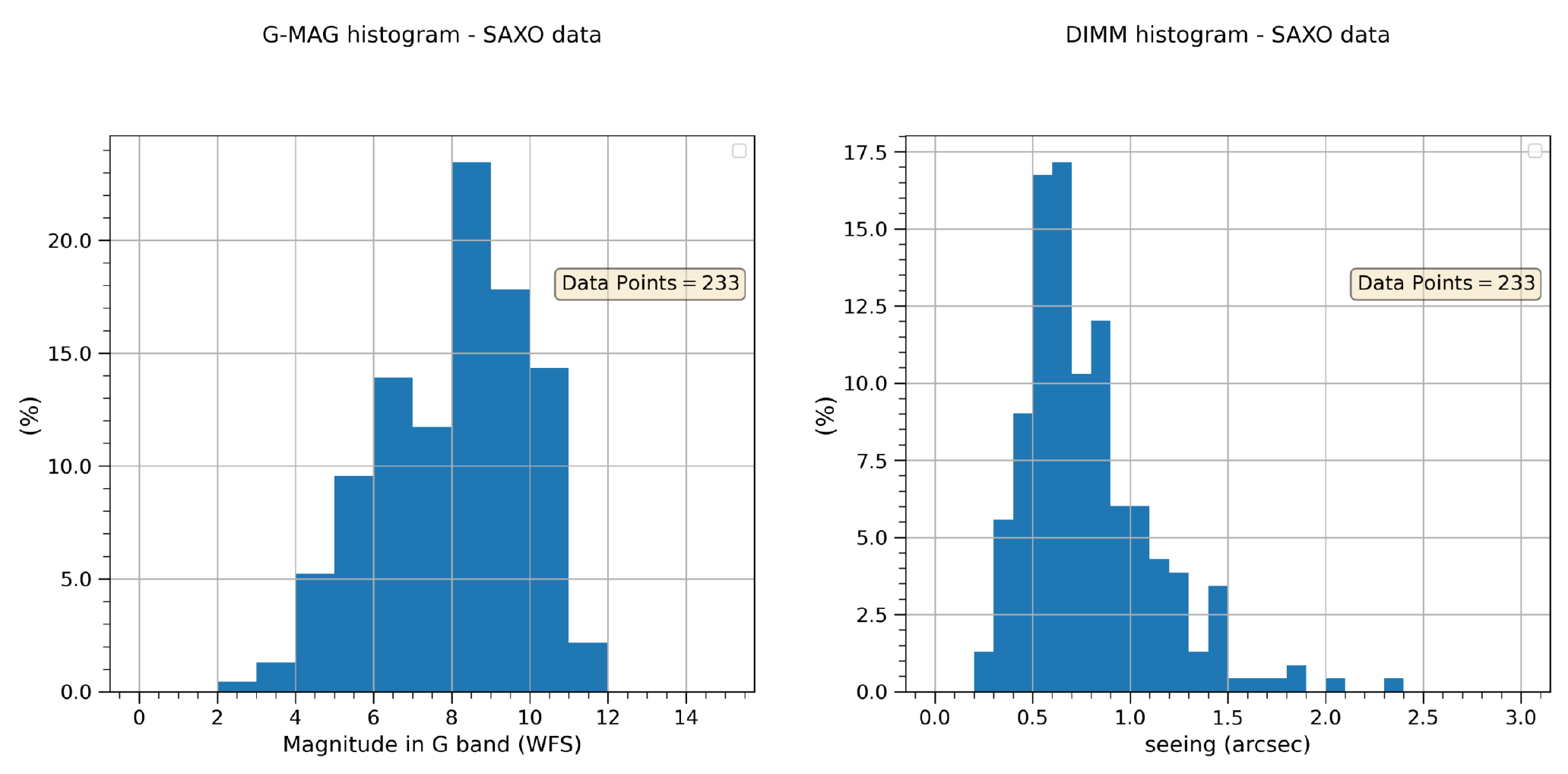}
\caption{Statistics of the SAXO measurement samples. Left: Distribution of the magnitude of the observed stars in G band (used for the simulations). Right: Distribution of the seeing values as measured by DIMM.}
\label{fig:saxomagplot}
\end{figure}

In order to validate the TIPTOP software, we compared its results with those obtained with PASSATA that we take as a reference. We performed this comparison using a larger PSF measurement database provided by the SAXO AO instrument, working with SPHERE at VLT. This sample is characterized by a completely different magnitude range with respect to SOUL, since SAXO (working with a planet finder) operates on much brighter magnitudes in the range [4-12], as shown in figure \ref{fig:saxomagplot}. Also seeing conditions at VLT are much better than at LBT, as shown by the DIMM measurements distribution over the dataset shown in the same figure. The FWHM and SR curves, with respect to magnitude, have different behaviours in the different magnitude ranges, with a larger FWHM dispersion to be expected for faint magnitudes, as shown in our previous study [\citenum{turchi2020s}]. Specifically the FWHM have a dependency on seeing that grows exponentially with the magnitude growing fainter. Having a dataset with bright magnitudes can allow us to complement the SOUL dataset that falls mainly with faint magnitudes.\\

\begin{figure}
\centering
\includegraphics[width=1.\textwidth]{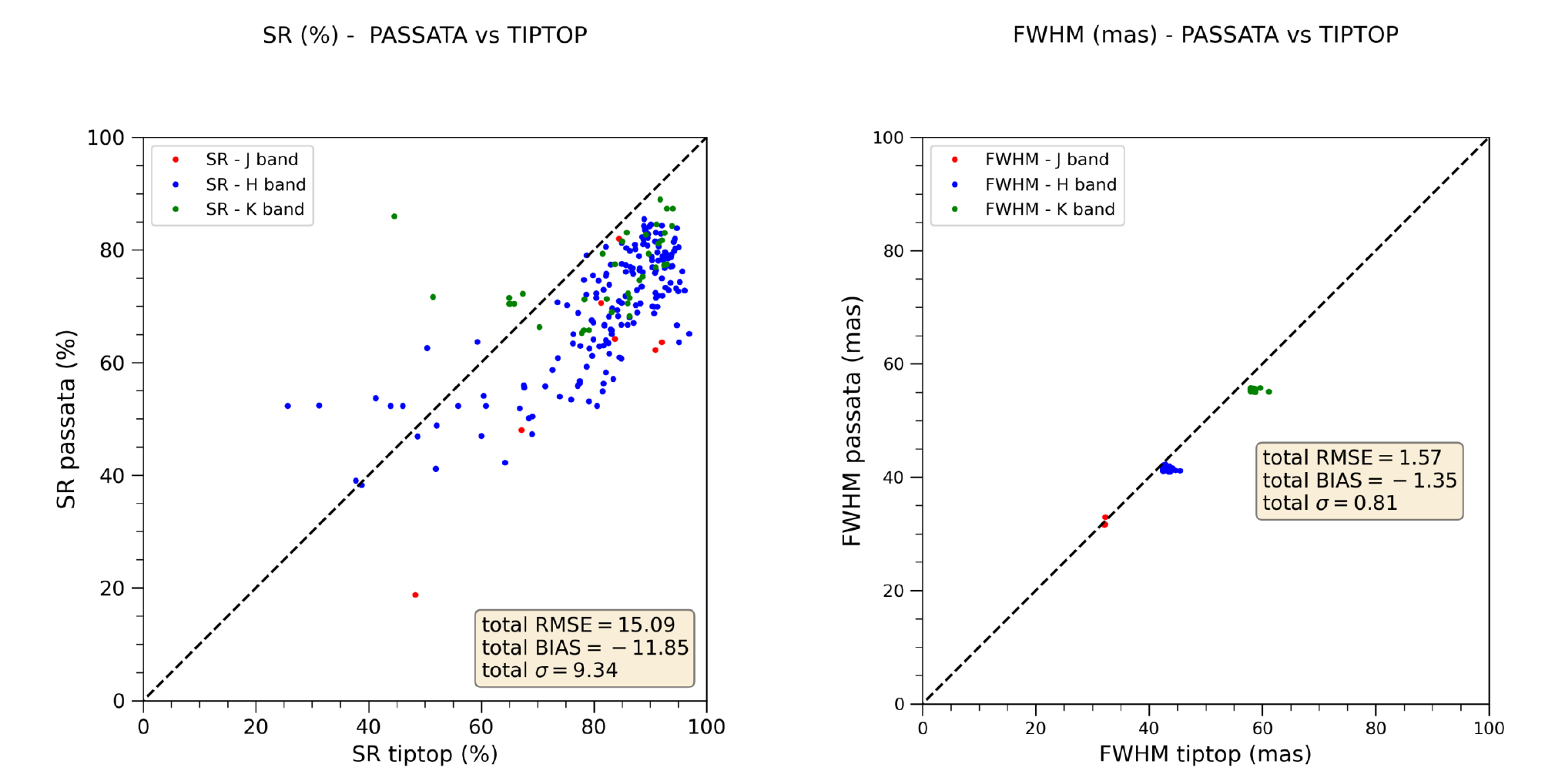}
\caption{Scatterplot of the results of the comparison of TIPTOP and PASSATA simulations for the SAXO case. Left: comparison of simulated SR. Right: Comparison of simulated FWHM. In each figure we report the total RMSE, BIAS and $\sigma$ obtained for the whole sample, while with different colors we plot the three observed bands (J, H and K).}
\label{fig:saxoscatterplot}
\end{figure}

\begin{figure}
\centering
\includegraphics[width=1.\textwidth]{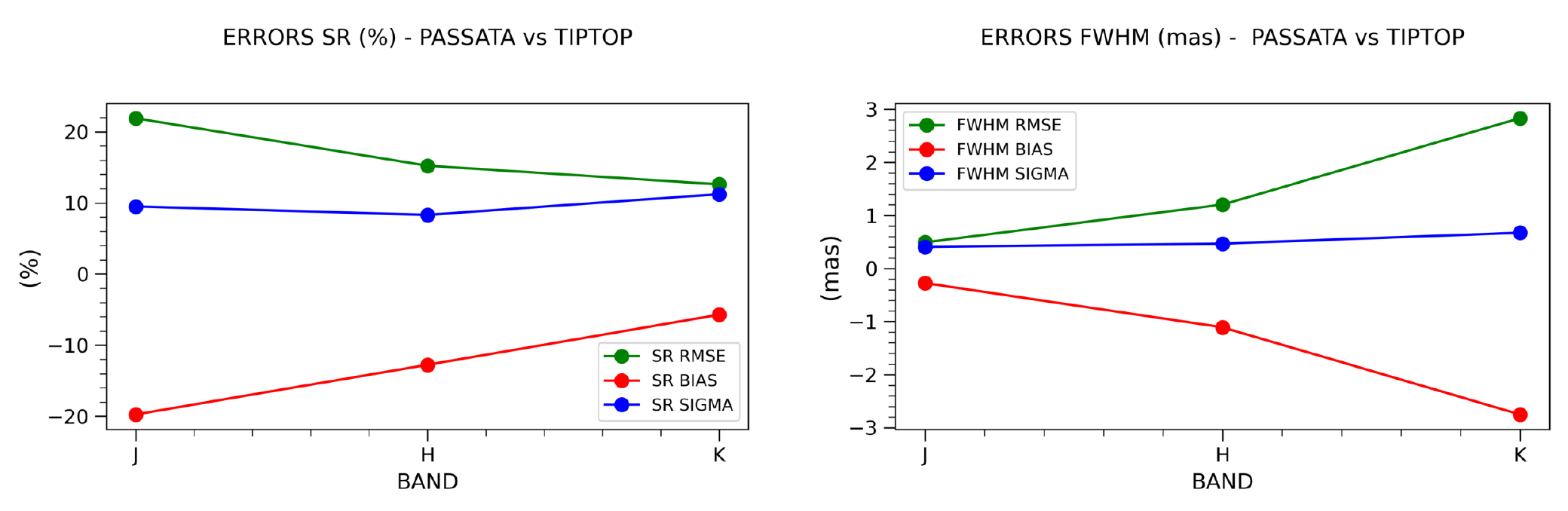}
\caption{Errors reported for each observed band in the comparison of simulations with TIPTOP and PASSATA. We report the RMSE, the BIAS and the SIGMA. Left: Errors on SR. Right: Errors on FWHM.}
\label{fig:saxoerrorplot}
\end{figure}

At present we are still working on the PSF measurements because there is the need to clean up the sample and provide the more accurate estimate for SR and FWHM, however we decided to start comparing the two PSF simulation softwares, PASSATA and TIPTOP, on the input data available for the SAXO PSF dataset, which is richer than the SOUL telemetry.\\
For PASSATA simulations we use as an input the seeing measured by the DIMM, the magnitude in G band (the one used by SAXO WFS), the WFS wavelength, the WFS frame rate and spectral bandwidth, the WFS COG window radius, the loop gain and delay (in frames). For TIPTOP we use the same inputs with the addition of the pixel scale and instead of the magnitude the number of photons per aperture per frame (NPH) on the WFS. In the PASSATA case since we have the gain parameter from the telemetry, we do not use the same linear fit method used in the SOUL case, thus we run a single simulation per run with the specific magnitude and a fixed timestep of 4s, however we still perform three runs (short simulated time) and average out the results (as wrote in Sec.~\ref{sec:soul} to produce a statistically reliable estimate). In the TIPTOP case, since it is analytic and compute directly long exposure PSF, we run a single simulation for each input set.\\
In Fig. \ref{fig:saxoscatterplot} and \ref{fig:saxoerrorplot} we report the results of this comparison.

From Fig. \ref{fig:saxoscatterplot} we immediately realize that the two simulation software matches perfectly for the FWHM case, in all three observed bands J, H, K. This is also due to the fact that, in the bright magnitude range, FWHM changes little with the different seeing and magnitude and no vibrations are considered here, so results tend to depend only on the observed band, as also evidenced in [\cite{turchi2020s}]. The SR comparison, while showing a good correlation between the two outputs, show a bias of around $\sim10\%$ which can be eventually corrected in postprocessing after comparison with the PSF measurements. However there is still a 10\% difference between the two software ($\sigma$, that is the pure error with subtracted bias) that can't be recovered from a bias correction TIPTOP is still in heavy development phase, so we will investigate to understand if this is a relevant error or is simply due to the different computational methods, approximations and model implemented in PASSATA and TIPTOP. In any case there is the need to evaluate the performances of the two software with respect to the PSF measurements, as soon as they will be available.\\

\section{Conclusions}
\label{sec:final}
In this paper we present the prosecution of our evaluation of the feasibility of a PSF operational forecast system, starting from our previous work [\cite{turchi2020s}]. By fine tuning the PASSATA software we already obtained a very promising result for the SOUL case, with an estimated accuracy on the SR forecast of order [10-15\%] depending on the observed band. The FWHM prediction, in the SOUL case, have a larger error, however it is probably due to the fact that we are operating in a very faint magnitude regime, where the AO correction is very sensible to small variations of seeing, magnitude, windshake and vibrations, as already reported in [\cite{turchi2020s}]. The major error contributions come from noise effects produced by vibrations and windshake, which are not included in the AO simulation. Also a very important aspect is to characterize the accuracy of FWHM and SR measurements obtained by the instrument telemetry. Future developments will move to consider the SAXO PSF dataset, which provides us a much richer sample with bright magnitudes, thus complementing the previous study performed on SOUL. Work is undergoing to clear the dataset and compute the PSF parameters.\\
In order to investigate the feasibility of a real-time operational tool we moved to evaluate the performance of TIPTOP software, which allows very fast computation times and could be the perfect candidate for satisfying the requirements of an operational tool. The comparison of the two software, performed on the inputs coming from the preliminary SAXO dataset, show a perfect agreement in the FWHM, although we have to remark that we are operating in a regime of bright magnitudes, thus making the FWHM weakly dependent from seeing and magnitude. The SR comparison show a difference of order $\sim10\%$ (after correction of the bias) from the two software, pointing out the need of further investigation, however providing us with an already acceptable confidence margin. Future developments will involve the setup of TIPTOP simulations also for the SOUL case, in order to investigate the faint magnitude case, and the evaluation of the performances of the two software on the SAXO PSF dataset.\\
It is interesting to note that the forecast of the seeing for ALTA short-term predictions on a 1-hour time frame is already indistinguishable from the DIMM measurements with an RMS of 0.1'', while the dispersion between DIMM and another instrument is of order 0.2'' [\citenum{masciadri2020,masciadri2022}]. So we expect that forecast error will not play a relevant role on the total error in the SR and FWHM prediction. This will be further studied when we will finally assess the PSF prediction strategy.\\
Our project, as a whole, will also move to evaluate off-axis cases by making use of the the full $C_N^2$ and wind stratification provided by Astro-Meso-NH model, which is the huge advantage provided by using an atmospheric OT forecasting tool. This will allow to extend the analysis and evaluate the impact in the case of Wide Field AO systems (WFAO), which will be the protagonists of the new generation of extremely large telescopes under construction.
\acknowledgments 
Study co-funded by the FCRF foundation action - N. 45103, ALTA Center (ENV002), Horizon 2020, Research and Innovation, SOLARNET (N. 8241135).


\end{document}